\pdfoutput=1

\documentclass[11pt]{article}

\usepackage[]{emnlp2021}

\usepackage{times}
\usepackage{latexsym}

\usepackage[ruled,vlined]{algorithm2e}
\usepackage{booktabs}
\usepackage{graphicx}
\usepackage{amssymb}

\AtBeginDocument{%
  \providecommand\BibTeX{{%
    \normalfont B\kern-0.5em{\scshape i\kern-0.25em b}\kern-0.8em\TeX}}}

\begin{document}

\title{Distilling Transformers for Neural Cross-Domain Search}

\author{Colin B. Clement\and Chen Wu\and Dawn Drain\and Neel Sundaresan\\
    Microsoft\\
    One Microsoft Way\\
    \texttt{\{colin.clement,chen.wu,dawn.drain,neels\}@microsoft.com}
}

\maketitle

\begin{abstract}

Pre-trained transformers have recently clinched top spots in the gamut of natural language tasks and pioneered solutions to software engineering tasks. Even information retrieval has not been immune to the charm of the transformer, though their large size and cost is generally a barrier to deployment. While there has been much work in streamlining, caching, and modifying transformer architectures for production, here we explore a new direction: distilling a large pre-trained translation model into a lightweight bi-encoder which can be efficiently cached and queried. We argue from a probabilistic perspective that sequence-to-sequence models are a conceptually ideal---albeit highly impractical---retriever. We derive a new distillation objective, implementing it as a data augmentation scheme. Using natural language source code search as a case study for cross-domain search, we demonstrate the validity of this idea by significantly improving upon the current leader of the CodeSearchNet challenge, a recent natural language code search benchmark.

\end{abstract}

\section{Introduction}

\newcommand{\codesearchnet}{\textsc{CodeSearchNet}~}

Pre-trained transformers~\cite{vaswani2017attention} have pushed the limits of natural language processing (NLP) tasks in both the domains of natural language and software engineering~\cite{clement2020pymt5,svyatkovskiy2020intellicode,tufano2020unit,guo2020graphcodebert}. Recent work using transformers for information retrieval tasks includes re-ranking~\cite{matsubara2020reranking}, caching document-level representations for efficient queries~\cite{macavaney2020efficient}, novel architectures like ColBERT~\cite{khattab2020colbert} which allows query-document interactions while caching embeddings for efficient queries and MarkedBERT~\cite{boualili2020markedbert} which merges traditional information retrieval (IR) cues into a BERT model. Here we approach cross-domain information retrieval from a novel angle by distilling~\cite{hinton2015distilling} expensive pre-trained sequence-to-sequence transformers into a lightweight bi-encoder which supports feature caching and efficient nearest neighbor searches.

Natural language code search takes in a natural language query, e.g. `perform an HTTPS POST', and retrieves a ranked set of code snippets which are most relevant to the intent of the query. The codes may share $n$-grams with the query, but the different domains of natural language (NL) and programming languages (PL) means this is not guaranteed. Further, there is an extra challenge as the available training data for code search---method-docstring pairs---is not precisely the domain of natural language queries and relevant codes. For this study we will use  \codesearchnet~\cite{husain2019codesearchnet}, a natural language code search benchmark for six programming languages (Python, Java, Javascript, Ruby, PHP, and Go). While other cross-domain IR tasks (e.g. English queries of German documents) implement some kind of cross-lingual token embedding~\cite{bonab2020training}, the \codesearchnet challenge shows that even standard baselines like ElasticSearch are competitive with large BERT-style encoders, and the top performing model at the time of writing is a Neural Bag of Words (NBoW) model.

This work will be concerned with bi-encoder code-search algorithms~\cite{gu2018deep,cambronero2019deep} which jointly embed natural language queries $q_i$ and code snippets $c_j$ in a continuous vector space in which queries are `close' to relevant code snippets. These embeddings are mapped with models like a Bag of Words, Neural Bag of Words, or a transformer encoder which ingests tokens representing some query $q_i$ and some code snippet $c_j$, and returns vector embeddings $E_q(q_i)$ and $E_c(c_i)$. $E_q$ and $E_c$ can be the same model, though we allow different embedding models for queries and codes.

These embeddings enable a `measure' of similarity, the most common being the dot product or cosine similarity. For some collection of queries and codes $\{q_i\}$ and $\{c_j\}$, where at training time it is known that, $c_i$ is `relevant' to $q_i$, the goal of the model is to encode them both so they are most `similar.' The empirical similarity is simply binary: the natural language query $q_i$ is a docstring and the relevant code $c_i$ is the snippet with which the docstring is naturally associated. The objective is to maximize the probability of these data conditioned on the parameters $\theta$ of a given model
\begin{eqnarray}
    \small
    \mathcal{L}(\theta) =& \log P(\{q_i\},\{c_j\}|\theta)\\
    =& - \sum_i \log q_\theta(q_i, c_i),\nonumber
    \label{eqn:max-likelihood}
\end{eqnarray}
where $N$ is the number of query-code pairs $q_i$, $c_i$, and
\begin{equation}\small
    \log q_\theta(q_i, c_j) \propto \frac{E_q(q_i, \theta)\cdot E_c(c_i, \theta)}{|E_q(q_i, \theta)||E_c(c_i,\theta)|}
    \label{eqn:bi-encoder}
\end{equation}
is a bi-encoder model of the log-likelihood. This objective encourages the embeddings of `matched' queries and codes $q_i$ and $c_i$ to be similar, and encourages every other `unmatched' pair of embeddings to be dissimilar (as the probabilities are normalized).

The  models are not encouraged to learn a spectrum of similarity for codes and queries because all other codes and queries are not exactly `matched.' Two functions with nearly identical behavior and similar docstrings will be pushed apart from one another in the embedding space, except for any helpful inductive biases in the embedding. Even a bi-encoder model with very intelligent inductive biases will be limited by the fact that the encoding of the query and code must be independent of one another for the retrieval to be efficient. Therefore, while the model will may high precision in finding a few codes it was trained to see, it will have poor recall in that it was trained to place all other code snippets not exactly matched to be `farther' away. Thus metrics like Normalized Discounted Cumulative Gain (NDCG) or Mean Reciprocal Rank (MRR) will suffer.

\section{Distilling translation models into bi-encoders}

The most powerful joint and conditional probability models currently available for sequences are transformer models~\cite{vaswani2017attention}, and have been successfully trained on bi-modal natural language code corpora~\cite{clement2020pymt5,svyatkovskiy2020intellicode,feng2020codebert}. \citet{penha2021calibration} found ensembles of BERT models and \citet{wang2020inference} found translation models to be well-calibrated. The models are woefully inefficient for practical retrieval; for each query one must score every single code in the corpus. How can we have the best of both worlds: a model which can powerfully generalize like a transformer while also admitting efficient queries? We train a transformer model to generally translate from queries to codes (slow to query) and use it to score or teach a bi-encoder to retrieve codes (fast to query), ideally preserving the generalized benefits of a powerful transformer model.

We use a teacher model $p(c_i|q_j)$ called PyMT5~\cite{clement2020pymt5} and assume all codes and queries are equally likely;  the prior probability $p(q_i)\sim p(c_j)\sim 1$. The original distillation paper~\cite{hinton2015distilling} studied the special case of matching the logits of the student and teacher model, which has convenient mathematical simplifications. Our student and teacher models have logits with incompatible dimensions (the teacher has a logit for every token, the student for each query-code pair). We proceed by using a metric to quantify the difference between student $q$ and teacher $p$, the Kullback-Leibler (KL) divergence~\cite{jaynes1957information}:
\begin{equation}
    \small
    \mathcal{D}_{KL}(p||q) = \sum_x p(x) \log\frac{p(x)}{q(x)}.
    \label{eqn:dkl}
\end{equation}
Note that if the teacher distribution is the empirical distribution of our data, i.e. $p(x)=1$ for $x$ observed in our data and $p(x)=0$ otherwise, this objective reduces to the original maximum likelihood objective of Eqn.~\ref{eqn:max-likelihood}.



Because Eqn.~\ref{eqn:dkl} is an expectation value over a teacher distribution of code-query pairs, we can express the objective as a sum of samples from the teacher
\begin{equation}
    \small
    \mathcal{L}(\theta) = -\frac{1}{NM}\sum_{(q,c)\sim p(c, q)}\log q_\theta(c, q),
    \label{eqn:sampled-objective}
\end{equation}
where $M$ is the number of code-query pairs sampled from the teacher $p(q, c)$. Here we use the same bi-encoder model as in Eqn.~\ref{eqn:bi-encoder} for $q_\theta(c, q)$. Now we can proceed, and in fact the interpretation of our new objective in Eqn.~\ref{eqn:sampled-objective} is quite simple: we can augment our existing data $\{q_i, c_i\}$, by adding $M$ sampled code-query pairs. In experiments that follow we will include the `matched' code as well, and not rely solely on the samples, as the data is assumed to derive from some desired distribution we are modeling.

\begin{figure}
    \centering
    \includegraphics[width=0.9\columnwidth]{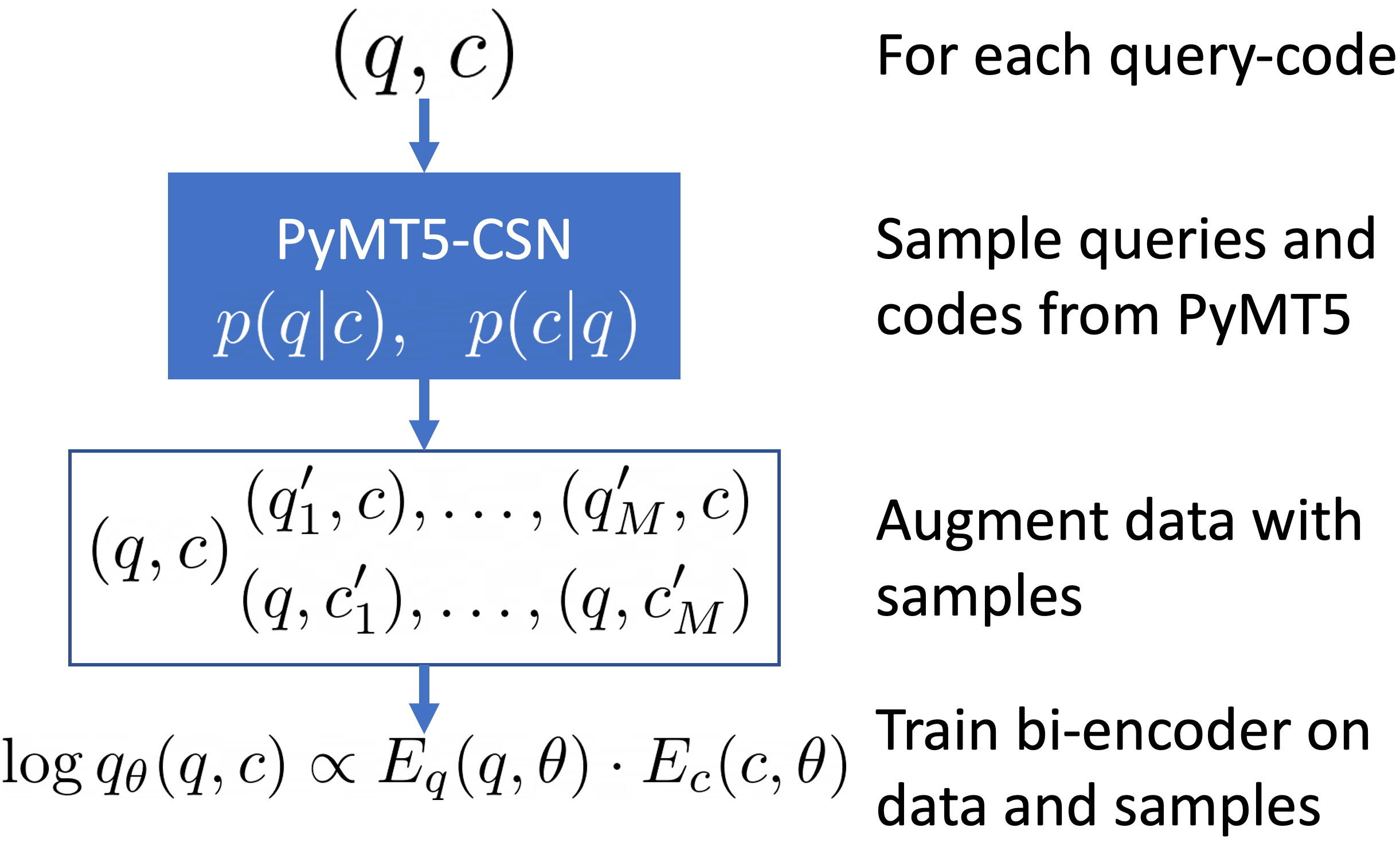}
    \caption{Architecture diagram of distillation procedure. For each query-code pair in our data set $(q, c)$, we 
    use a PyMT5 fine-tuned on the same training set to sample $M$ queries and $M$ codes conditioned on the true code and query, respectively. We then combine this sampled data with the original data to train the bi-encoder retrieval model.}
    \label{fig:distill}
\end{figure}

Finally, complying with the manifest query-code symmetry of the bi-encoder joint probability, the distillation objective we propose is (up to a multiplicative constant)
\begin{equation}
    \resizebox{\columnwidth}{!}{
    $\displaystyle{\mathcal{L}(\theta) = -\sum_i \left(
        \sum_{c\sim p(c|q_i)}\log q_\theta(c, q_i) + \sum_{q\sim p(q|c_i)}\log q_\theta(c_i, q)
    \right).
    }$
    }
    \label{eqn:sampled-objective}
\end{equation}
We implement the minimization of this objective by augmenting the training data with $M$ sampled code and docstrings, shuffling them all together, and using stochastic gradient descent. Figure~\ref{fig:distill} shows this process, whereby for each true query-code pair, we sample $M$ queries and $M$ codes from a teacher model, conditioned on the true code and query, respectively. That is, for $M=1$, we have tripled the size of our data pairs. We then feed this augmented data into a standard mini-batch stochastic gradient descent trainer to distill the teacher model into the bi-encoder. Careful readers may note this algorithm is similar in character to \citet{nogueira2019document} and \citet{ma2020zero}.

\begin{table*}[htbp]
    \centering\small
    \begin{tabular}{llllllllll}
                                    & Python    & Java      & Ruby      & JS        & PHP           & Go            & All & Valid. NLL\\\midrule
        CSN-PyMT5 teacher           & 0.651     & 0.609     & 0.525     & 0.531     & 0.514         & 0.701         & 0.665 & 1.008 \\\midrule
        hybrid-NBoW baseline        & 0.614     & 0.553     & 0.447     & 0.491     & 0.508         & 0.676         & 0.646 & 1.020\\
        hybrid-NBoW distill M=1   & \bf0.662  & 0.602     & 0.515     & 0.526     & 0.522         & 0.697         & 0.667 & 1.008\\
        hybrid-NBoW distill M=2   & 0.661     & \bf 0.609 & \bf 0.525 & 0.531     & \bf 0.524     & \bf 0.701     & \bf 0.672 & \bf1.006\\
        hybrid-NBoW distill M=4   & 0.646     &  0.603    &  0.522    & \bf 0.533 & 0.512         & 0.700         &   0.660 & 1.009
    \end{tabular}
    \caption{Mean reciprocal rank (MRR) and negative log-likelihood (NLL) scores for each PL language and all languages, evaluated on the CSN test set for a hybrid-NBoW model baseline and distilled hybrid-NBoW with $M$ teacher samples. }
    \label{tab:mrr-results}
\end{table*}

\subsection{Teacher Model CSN-PyMT5}

We used a 400 million parameter version of PyMT5, trained on 26 million Python methods (7.7 million of which possessed docstrings), fine-tuned on the \codesearchnet. The model converged after 10 epochs, about 8 hours on 16 Tesla V100 32GB GPUs. Using the same CSN training set, we then sampled $M=10$ docstrings for each method and $M=10$ methods for each docstring.

\subsection{Student Model}

The model currently on the top of the CSN challenge leaderboard is a hybrid-NBoW model  (though CodeBERT has reported superior MRR results~\cite{feng2020codebert}). 
The hybrid-NBoW model uses three separate reduction methods on the vector sequence---max-pooling, averaging, and self-attention and further learns a weighted average of these three reductions. 
We allowed a separate encoder for queries and each PL, with a total of 7 encoders, and about 9 million total parameters (or 1/40 as large as the teacher model PyMT5). We trained the students and the baseline using the default NBoW hyperparameters in the \codesearchnet library.

\section{Distillation Experiments}

Table~\ref{tab:mrr-results} shows the results of several experiments. The first row is the teacher model CSN-PyMT5, which was evaluated by inefficiently scoring every single pair of queries and codes in the test set (100 million pairs) to find the rank of the correct code. Below that is the baseline and current leader of the \codesearchnet challenge, which is beat by the teacher. The second row is the same model trained with one ($M=1$) sampled docstring and method for each real training pair. A consistent improvement for all languages can be seen, especially for Python, which was the language with which the teacher model PyMT5 was first trained. Interestingly, Ruby shows the greatest absolute improvement. Ruby which shares syntactic features with python like the \texttt{def} keyword and indentation for scope information.

For two samples, the third row of Tab.~\ref{tab:mrr-results}, further improvement for all languages except Python can be seen. The regression of Python compared to $M=1$ is perhaps insignificant, but there are clearly diminishing gains as the absolute improvement from $M=1$ is less than between the baseline and $M=1$. At four samples ($M=4$) in the fourth row of Tab.~\ref{tab:mrr-results}, we see a complete regression, where overall and for all languages except JavaScript the $M=2$ distillation is superior. Perhaps most interestingly, the student can beat the teacher both in MRR and validation loss.

Recall that \codesearchnet aims to train a model to retrieve code methods by training on matched natural language docstring and code method pairs. Naturally, a person searching for code would likely not type out a docstring query, so CSN provides 99 real search-engine queries with an evaluation set of code methods which have been hand-labeled by programming experts. Table~\ref{tab:ndcg-results} shows the result of this evaluation, comparing our baseline with an $M=1$ distillation. Our distilled model is beaten across the board. This is perhaps not that surprising, as we have used a powerful teacher model adept in the distribution of docstrings and methods, thus it follows our student has the same bias.

\begin{table}[ht]
    \centering\small
    \resizebox{\columnwidth}{!}{
    \begin{tabular}{lllllllll}
                  & Python    & Java      & Ruby      & JS        & PHP           & Go            & All\\\midrule
        Teacher   & 0.46  & 0.40  & 0.36  & 0.34  & 0.32      & \bf 0.35      & 0.37 \\\midrule
        baseline  & \bf0.47  & \bf0.43  & \bf0.38  & \bf0.36  & 0.34      & \bf0.33      & \bf0.38\\
        M=1   & 0.43     & 0.38     & 0.35     & 0.32     & 0.30         & 0.29         & 0.35\\
    \end{tabular}
    }
    \caption{Normalized Discounted Cumulative Gain (NDCG) scores on 99 search-engine queries hand-ranked by expert programmers. The search engine queries are distributed differently than source code docstrings. The student and baseline models are the same hybrid-NBoW used throughout this work.}
    \label{tab:ndcg-results}
\end{table}

\section{Discussion}



\citet{penha2021calibration} studied in depth the calibration of BERT-style models for IR tasks. This area should be explored further. In particular, one should investigate the assumptions we have made, that translation models are excellent retrievers. For example, one could retrieve with ElasticSearch, and use the translation model to re-rank the retrieved results as a baseline. For \codesearchnet ElasticSearch was a strong baseline; this approach could potentially approximate an upper bound of the student model performance. If this line of inquiry finds that translation models are not well-calibrated, one could apply scalar or even vector Platt scaling following \citet{wang2020inference}.

The unique issue of \codesearchnet is the misalignment between real queries and available natural language docstrings. Our experiments show that distillation can reinforce the difference. Some natural question are: can we create a teacher which is adept also at modeling real queries? Would the distillation process bias the student---even using docstring data for training---to perform better with real queries? One approach might be to obtain a corpus of real search queries filtered by code-search intent and a corpus of codes (e.g. code bodies in Stack Exchange web pages) and use an unsupervised translation approach~\cite{lachaux2020unsupervised}. If real search results are available, pursuing an effort to incorporate user choices or feedback could be even more meaningful~\cite{zamani2020analyzing}. 

\section{Conclusion}

We introduced a novel method to distill a powerful translation model---which is infeasible to use for information retrieval---into a lightweight, easily queried bi-encoder 1/40th the size. We demonstrated our approach by improving the top model on the \codesearchnet challenge leaderboard. Our method, derived from first principles, is easy to implement through a data augmentation scheme wherein one fine-tunes a pre-trained translation model on query-document pairs, drawing sampled queries and documents from the real pairs. We found diminishing returns with 2 samples outperforming 4 samples. We also found the distilled model could beat the teacher, and that distillation reinforces the training data distribution, causing our model to perform worse on retrieving codes from real search queries instead of the code docstrings it was trained on. While the original method of distillation matches logits, our method is  general and can be used on any student/teacher architectures. Our use emphasizes that distillation can not only reduce model sizes, but also improve the algorithmic complexity of inference.

\section*{Acknowledgments}
The authors would like to thank Alex Alemi for helpful direction, and Alexey Svyatkovskiy and Michele Tufano for feedback on the manuscript, and Liz Baker for copy-editing.

\bibliographystyle{ACM-Reference-Format}
\bibliography{codesearch,local}

\end{document}